\documentclass[twocolumn, twocolappendix, times]{aastex631}
\usepackage{txfonts}

\accepted{22/Jul/2026}

\graphicspath{{./}{figs}}

\begin{document}
\title{When galaxies cross cold fronts: wind tunnel simulations of ram pressure stripping}

\author[0009-0007-1135-6474]{Elvis A. Mello-Terencio}
\email{elvisamandio@on.br}
\affiliation{Departamento Acadêmico de Física, Universidade Tecnológica Federal do Paraná, Av. Sete de Setembro 3165, Curitiba, PR, Brazil}
\affiliation{Observatório Nacional, Rua Gal. José Cristino 77, Rio de Janeiro, RJ, Brazil}

\author[0000-0001-7319-297X]{Rubens E. G. Machado}
\affiliation{Departamento Acadêmico de Física, Universidade Tecnológica Federal do Paraná, Av. Sete de Setembro 3165, Curitiba, PR, Brazil}
\affiliation{Instituto de Astronomia, Geof\'isica e Ci\^encias Atmosf\'ericas, Universidade de S\~ao Paulo, Rua do Mat\~ao 1226, S\~ao Paulo, SP, Brazil}

\author[0009-0006-9462-1044]{Richards P. Albuquerque}
\affiliation{Instituto de Astronomia, Geof\'isica e Ci\^encias Atmosf\'ericas, Universidade de S\~ao Paulo, Rua do Mat\~ao 1226, S\~ao Paulo, SP, Brazil}

\begin{abstract}

Cluster collisions and mergers are among the most energetic phenomena in the low redshift Universe. These interactions disturb the intracluster medium creating regions with density and temperature discontinuities, such as sloshing spirals. There is evidence that such environments can influence galaxy evolution. This study aims to understand how a galaxy that crosses an environment with discontinuities in density and temperature can be affected. To this end, a set of simulations was conducted using a wind tunnel setup into which a {MW-like mass} galaxy was inserted. A total of eight models were created, namely two control runs and six with distinct density and temperature transitions along the tunnel, comprising simulations with low density and high density environments. Results show that galaxies lose {considerably} more gas due to the higher density encountered when crossing denser and discontinuous regions in comparison to a constant density environment. The star formation rate exhibits a brief enhancement when the galaxy enters the denser section of the tunnel and {(u-i)} color index also undergoes slight changes, initially becoming bluer. However, even in the simulations with the most intense transitions, changes {in star formation rate and color index} are not substantial, reaching at most 5\% difference in relation to the control models. By the end of the simulation runs, star formation rate and color index are similar to the control runs. These results suggest that crossing intracluster medium discontinuities can induce measurable effects in a galaxy, but these are subtle and short-lived.

\end{abstract}



\section{Introduction}

As disc galaxies traverse the intracluster medium (ICM), their gas content will be partially removed by ram pressure stripping \citep{Gunn1972}. Not only can this mechanism remove gas from the interstellar medium (ISM), it can also deplete the gaseous halo of the galaxy from gas preventing future gas infall and contributing to star formation quenching \citep{Larson1980, Balogh2000}. In extreme cases of ram pressure stripping, the galaxy develops a ``jellyfish'' tail, composed mainly of gas filaments or clumps extending opposite from the galaxy's direction of motion. This interaction between the ICM and ISM is one of the major environmental influences that can alter the evolution of galaxies. For a recent review, see \cite{Boselli2022}.

Ram pressure stripping has also been the subject of numerous studies using hydrodynamical simulations. These analyses have further confirmed that ram pressure is the main process by which galaxies lose gas \citep{Roediger2005, Jachym2007, Tonnesen2007} and may be affected by many parameters such as mach number \citep{Roediger2005}, density of the region crossed and relative velocity \citep{Tonnesen2008, Tonnesen2011, Steinhauser2012}, the specific trajectory through the host galaxy cluster \citep{Tonnesen2009, Ruggiero2017, Tonnesen2019}, and the inclination of the galaxy plane relative to the wind \citep{Roediger2006, Kronberger2008, Jachym2009}. These simulations also show that ram pressure impacts star formation rate (SFR), either quenching it \citep{Steinhauser2016, Lee2020} or enhancing it \citep{Kapferer2008, Kronberger2008} depending on the ICM properties. Furthermore, naturally occurring jellyfish galaxy populations in cosmological simulations have also shown correlation between their motion, morphologies and the time scale they are observable \citep{Yun2019, Joshi2020, Troncoso2020}. More recently, studies on individual galaxies have also been conducted, focusing on reproducing the exact behavior caused by ram pressure stripping on observed galaxies \citep{Vollmer2021} or modeling more specific aspects, such as the interaction of ram pressure with black holes \citep{Akerman2023}.

The presence of cold fronts is common in the centers of relaxed clusters with cool cores, where smaller interactions can easily upset the gas and create sloshing spirals \citep{Ascasibar2006}. Mergers and close encounters also create significant shock fronts at the intersection between clusters, where the ICM from each cluster interacts \citep{Markevitch2001, Ruggiero2019, Albuquerque2024}. Such gas regions are characterized by abrupt transitions of density and temperature, creating unique environments for a galaxy that may cross it. Since density is {one of} the main factors controlling ram pressure, it is expected that such transitions influence the evolution of galaxies. Observational results corroborate with this idea, where \cite{Owers2012} identified jellyfish galaxies in proximity to features associated with ICM merging shocks, \cite{McPartland2016} results suggest that extreme cases of ram pressure stripping are triggered by cluster mergers, and \cite{Churazov2023} discussed the scenario of a radio galaxy that appears to have recently interacted with a high Mach number shock front, possibly enhancing its emissions. Using cosmological simulations, \cite{Vijayaraghavan2013} found that merger shocks enhanced ram pressure, which significantly decreased the stripping radius of galaxies in a group. Moreover, \cite{RomanOliveira2019} studied the stripped galaxy population in the multicluster system Abell~901/2, finding that jellyfishes were preferentially located near the gas boundaries in between merging clusters. Within the same system, \cite{Ruggiero2019} applied hydrodynamical simulations to explore ICM properties, concluding that disturbed regions present in interacting clusters may induce the formation of more extreme jellyfish galaxies. More recently, \cite{Lourenco2023} found a weak correlation on a set of observable galaxies across 52 different clusters, where the fraction of ram pressure stripped galaxies is marginally higher in interacting clusters, and suggests that short-lived enhanced RPS may be the cause.

In this study, we aim to understand how crossing a discontinuous density environment such as those seen in cold fronts from interacting clusters may impact galaxy evolution. We employ $N$-body hydrodynamical simulations with the {\sc Arepo} code and model a galaxy within a periodic volume, effectively emulating a wind tunnel. We reproduce a galaxy crossing a sloshing spiral cold front by recreating the sharp ICM density and temperature discontinuities with idealized profiles along the simulation volume and compare the results with control runs considering a constant density tunnel. {Our analysis evaluates the evolution of galaxy morphology, gas content, star formation rate, and color index throughout simulations.}

This study is structured as follows: section \ref{sec:methods} describes the methods and setup procedures for creating initial conditions and running the simulations; section \ref{sec:results} presents the results obtained from the models; and in section \ref{sec:discussion} we present the discussion, relating the results obtained with current literature. At section \ref{sec:conclusions} we summarize all findings and conclude this study, suggesting future developments.


\section{Simulation Setup and methods}
\label{sec:methods}

{\color{gray}

}

In order to recreate the environment of a galaxy crossing sloshing spirals and study the effects of ram pressure stripping on a disc galaxy, $N$-body simulations were employed with the {\sc Arepo} moving mesh hydrodynamics code. {Star formation is modeled with the multiphase ISM model of \cite{Springel2003}. Gas above a certain density threshold is treated as a two-phase medium, composed of cold star-forming clouds and a hot ambient medium. Stellar population particles are spawned probabilistically depending on the local star formation rate}. {Supernova feedback is modeled by directly injecting thermal energy into the ambient hot phase following the depletion of cold gas clouds by star formation.} Our approach involved using a periodic gas box, effectively emulating a wind tunnel with idealized temperature and density profiles, where a galaxy would be placed and experience different physical properties when interacting with the simulated ICM.

The following subsections describe the simulation setup in detail, first explaining the environment of a cold front, then how the simulation box was created to reproduce its properties, followed by how the galaxy is created and inserted into this wind tunnel. A description of how galaxy quantities were measured is also given in the last subsection.

\subsection{Sloshing spirals}
\label{sec:sloshing_spirals}

Sloshing spirals are characterized by their distinctive morphology of cold gas along the ICM, forming spiral patterns with strong density and temperature gradients on the interface with the regular ICM.

Simulations show that the parameters that lead to sloshing spirals have relatively loose conditions, being easily set off by minor mergers \citep{Ascasibar2006}. These broad conditions allow for a multitude of combinations, providing flexibility to justify the simulation parameters chosen in the next sections.

\begin{figure}
    \centering
    \includegraphics{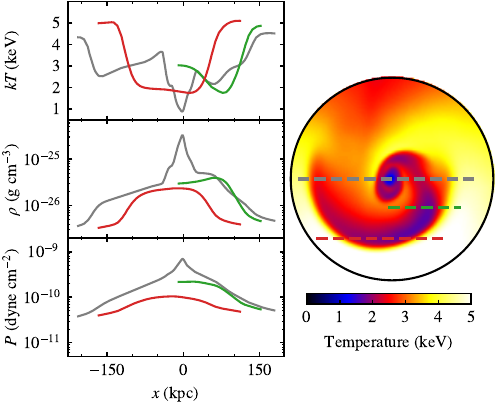}
    \caption{On the left are the profiles of temperature (top), density (middle), and thermodynamic pressure (bottom) corresponding to each trajectory shown in the temperature map. On the right is the temperature map of the sloshing phenomenon within a circular region 450 kpc in diameter. The profiles were measured along the directions represented by the dashed lines of the corresponding color.}
    \label{fig:sloshing}
\end{figure}

In order to motivate the typical orders of magnitude involved, we present here an example of a simulation of the collision between two idealized clusters where a sloshing spiral emerges, as illustrated in Fig.~\ref{fig:sloshing}. These clusters were generated using \cite{Hernquist1990} profiles for dark matter:
\begin{equation}
\label{eq:hernquist_profile}
\rho_{\rm h}(r) = \frac{M_{\rm h}}{2\pi} \frac{a_{\rm h}}{r(r+a_{\rm h})^3},
\end{equation}
where $M_{\rm h}$ is the total dark matter mass and $a_{\rm h}$ is the scale length. For the main (sub)cluster, we adopted a mass of $M_{200} = 5.1 \times 10^{14}\,{\rm M_\odot}$ ($M_{200} = 1.6 \times 10^{14}\,{\rm M_\odot}$), with $a_{\rm h} = 608$\,kpc ($a_{\rm h} = 387$\,kpc).

{The gas component was modeled with \cite{Dehnen1993} profiles of the form}
\begin{equation}
\label{eq:dehnen_profile}
\rho_{\rm g}(r) = \frac{3M_{\rm g}}{4\pi} \frac{a_{\rm g}}{(r+a_{\rm g})^{4}},
\end{equation}
with total mass $M_{\rm g} = 1.4\times10^{14}\,{\rm M}_{\odot}$ and scale length $a_{\rm g} = 494\,{\rm kpc}$ for the main cluster, and $M_{\rm g} = 4.4\times10^{13}\,{\rm M}_{\odot}$ and $a_{\rm g} = 335\,{\rm kpc}$ for the secondary cluster. At the pericentric passage these clusters were separated by 260\,kpc. Fig.~\ref{fig:sloshing} refers to a snapshot 1.14\,Gyr after the pericentric passage, when the sloshing spiral is well developed. In the temperature map on the right panel from Fig.~\ref{fig:sloshing}, a galaxy following the dashed red trajectory, for example, initially travels through a less dense and hotter environment, subsequently crossing temperature and density interfaces and transitioning into a denser and colder region. In this particular case, a galaxy would experience a temperature variation of $\sim 3\,{\rm keV}$ moving through a region 10 times denser than its initial environment. {The thermodynamic pressure curves in the lower plot of Fig.~\ref{fig:sloshing} serve as a way to probe the stability of the regions, as will be further discussed in the next section.}

The gray and green remaining dashed lines in Fig.~\ref{fig:sloshing} also show trajectories with similar variations, further illustrating the magnitude of these transitions in different scenarios. In the next subsection, initial conditions will be designed to recreate such environments in a more controlled and deliberate manner, motivated by the approximate ranges of values and ratios seen in this binary collision.

\subsection{Simulation box}

For all simulations we used a rectangular computational volume with dimensions $x\times y\times z = 100\,{\rm kpc}\times 100\,{\rm kpc}\times 200\,{\rm kpc}$ employing periodic boundaries in all directions. In order to recreate the transitions seen in sloshing spirals we then set gas properties along all the simulation volume emulating the temperature and density profiles along the $z$ direction following an analytical function.

Preliminary tests with juxtaposed sections of different densities and temperatures produced non physical pressure waves along the box. These early tests resulted in severely unstable initial conditions, due to numerical problems caused by excessively large gradients in the gas quantities.

Subsequent attempts at mitigating unwanted pressure waves aimed primarily at smoothing the transitions, forming pressure curves that more closely resembled the last plot of Fig.~\ref{fig:sloshing}. Linear profiles for these sections still provoked similar instabilities, probably caused by its characteristics at the beginning and end of each transition, where quantities were not as gentle. This motivated the introduction of a smooth analytical function for the required transitions that would keep the box stable and emulate the sloshing effect.



We settled on using transitions based on logistic curves, with a generic form given by:
\begin{equation}
    \label{eq:simple_logistic}
    f(z) = \frac{A}{1 + e^{-k(z-z_0)}} + B,
\end{equation}
with $A$ the maximum amplitude of the function, $k$ the steepness of the transition, $z_0$ the center of the function's midpoint and $B$ the minimum amplitude. This ensured a smooth transition between the low and high density sections.

\begin{figure}
    \centering
    \includegraphics[width=\columnwidth]{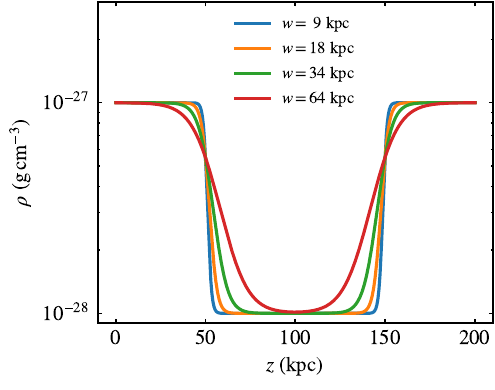}
    \caption{Examples of a density profile using the combined logistic function. The width of the transition region $w$ is the distance between the percentiles 1\% and 99\% of the curve amplitude.}
    \label{fig:function_comparison}
\end{figure}

The full equation used for distributing the gas box densities was a combination of two curves from equation \ref{eq:simple_logistic} as:
\begin{equation}
    \label{eq:double_logistic}
    g(z) = -\frac{\rho_{\rm{peak}} - \rho_{\rm{base}}}{1 + e^{-k(z-z_0)}} + \frac{\rho_{\rm{peak}} - \rho_{\rm{base}}}{1 + e^{-k(z-z_1)}} + \rho_{\rm{peak}}
\end{equation}
{where we considered for our setup $\rho_{\rm{base}}$ the amplitude of the curve between transitions, $\rho_{\rm{peak}}$ the density before and after transitions}, $k$ the steepness of those transitions, and $z_0$, $z_1$ the positions of the first and second transitions, respectively. For convenience, a width $w$ of the transition was deduced, based on the distance between points where equation \ref{eq:simple_logistic} reach percentiles $1-p$ and $p$ of the amplitude as:
\begin{equation}
    \label{eq:width_logistic}
    w = \frac{2}{k}\ln{\frac{p}{1-p}}.
\end{equation}
This can be shown to also hold true for transitions in equation~\ref{eq:double_logistic}. Fig.~\ref{fig:function_comparison} illustrates examples of density profiles with distinct values of the width $w$. The adopted transition width/steepness for all simulations was $w \approx 9.8\,{\rm kpc}$, which is equivalent to $k=0.94\,{\rm kpc^{-1}}$ using a percentile $p=0.99$. This value was chosen empirically by testing numerous boxes and selecting the largest percentile that still kept box stability and ensured a smooth and quick transition with our mass resolution. The transition middle points were set to $z_0 = 50\,{\rm kpc}$ and $z_1 = 150\,{\rm kpc}$.

After assigning the densities with equation \ref{eq:double_logistic}, the temperature was set at the center region to 2 keV in all boxes. The temperature along the remaining gas cells is such that the pressure is constant throughout the box, assuming an ideal gas. This ensures hydrostatic equilibrium, which keeps the created interfaces stable during simulations. {As opposed to the smooth and slightly varying curves on the bottom plot of Fig.~\ref{fig:sloshing}, if constant pressure is not enforced in these initial conditions as they were defined}, pressure gradients might arise and form waves propagating along the simulation box, which would be non physical.

For the resulting gas boxes used in our simulations, {two values of the base density $\rho_{\rm{base}}$in equation \ref{eq:double_logistic}} were adopted, namely LD (low density) with $\rho_{\rm{base}}=1\times 10^{-28}$\,g\,cm$^{-3}$ and HD (high density) with $\rho_{\rm{base}}=1\times 10^{-27}$\,g\,cm$^{-3}$. These are equivalent to electron number densities of $n=4.6\times 10^{-5}$\,cm$^{-3}$ and $n=4.6\times 10^{-4}$\,cm$^{-3}$, respectively. These specific values were chosen to represent typical regions of a cluster where a galaxy may encounter a cold front. Overall, these are comparable to gas densities found at radii of a few 100\,kpc of a cluster such as the one considered in section \ref{sec:sloshing_spirals}. 

\begin{figure}
    \centering
    \includegraphics{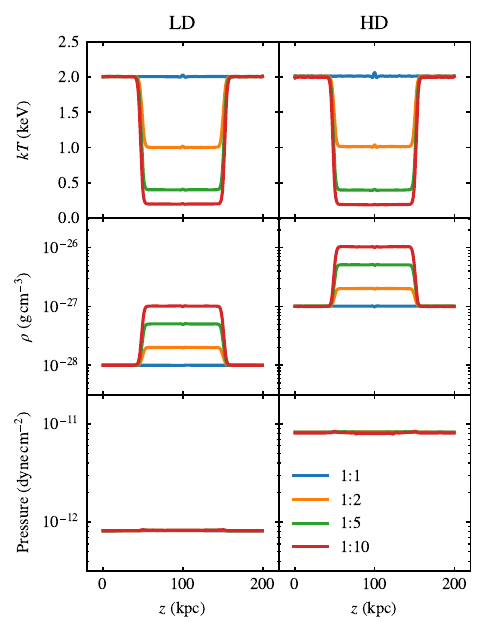}
    \caption{Temperature $kT$ (top), density $\rho$ (middle), and thermodynamic pressure (bottom) profiles for all gas boxes used at $t=1.1$\,Gyr, when the galaxy is inside the coldest and densest gas. Profiles for the LD models are shown on the left, while the HD profiles are on the right.}
    \label{fig:box_profiles}
\end{figure}

For each base density value, four different boxes were then created: a control box of uniform density, and three boxes where the amplitude of the discontinuity is given by factors of 1:2, 1:5 and 1:10 relative to the base density. The uniform boxes are labeled LD1 and HD1. The remaining boxes are labeled according to the amplitude of the transition: LD2 and HD2, LD5 and HD5, and LD10 and HD10. All these models are summarized in table \ref{tab:ic}, {along with the corresponding density parameters from equation\,\ref{eq:double_logistic}, namely $\rho_{\rm{base}}$, the base density, and $\rho_{\rm{peak}}$, the peak density of each box}. The profiles used can be seen in Fig.~\ref{fig:box_profiles}, which displays the simulations after the galaxy has traversed the box for $0.1$\,Gyr. We emphasize that the LD1 and HD1 models represent a galaxy traversing a uniform ICM, where no density interface is encountered and the gas distribution remains constant, as opposed to other models with varying interface amplitude.


This process for generating the ICM gas for the simulation box created stable conditions for all tests and simulations in the time scale needed for our application. The very minor spikes in density and temperature seen in Fig~\ref{fig:box_profiles} correspond to the galaxies inserted in those simulations while the minor pressure variations in the bottom most plots don't influence the simulations in any perceivable way.

\begin{table}
\caption{Names of the models, gas densities, and amplitude of the discontinuity used for each model}
\label{tab:ic}
\begin{center}
\begin{tabular}{l l l l}
\hline
Label & Base density & Peak density & Amplitude of\\
 & $\rho_{\rm{base}}$ $({\rm g}\,{\rm cm}^{-3})$ & $\rho_{\rm{peak}}$ $({\rm g}\,{\rm cm}^{-3})$ & transition \\
\hline
LD1     & $10^{-28}$ & $1\times10^{-28}$ & $1:1$ \\
LD2     & $10^{-28}$ & $2\times10^{-28}$ & $1:2$ \\
LD5     & $10^{-28}$ & $5\times10^{-28}$ & $1:5$ \\
LD10    & $10^{-28}$ & $1\times10^{-27}$ & $1:10$ \\[0.5em]
HD1     & $10^{-27}$ & $1\times10^{-27}$ & $1:1$ \\
HD2     & $10^{-27}$ & $2\times10^{-27}$ & $1:2$ \\
HD5     & $10^{-27}$ & $5\times10^{-27}$ & $1:5$ \\
HD10    & $10^{-27}$ & $1\times10^{-26}$ & $1:10$ \\
\hline
\end{tabular}
\end{center}
\end{table}

\subsection{Galaxy initial conditions}

The galaxy was created using {\sc Galstep} \citep{Ruggiero2017}, with the recent Python3 version\footnote{\href{https://github.com/elvismello/galstep}{https://github.com/elvismello/galstep}}. This tool creates an idealized galaxy composed of a dark matter halo, a bulge, a stellar disc, and a gas disc. The dark matter halo and the bulge follow a \cite{Hernquist1990} density profile with $M_{\rm h} = 10^{12}\,{\rm M}_{\odot}$ and $M_{\rm b} = 2\times10^{10}\,{\rm M}_{\odot}$ total mass, and $a_{\rm h} = 47$\,kpc  $a_{\rm b} = 1.5$\,kpc for their scale lengths, respectively.

The stellar and gas disc components are modeled with a double exponential profile:
\begin{equation}
\label{eq:double_exponential_profile}
\rho_{\rm d}(R,z ) = \frac{M_{\rm d}}{4\pi R_{\rm d}^2 z_{\rm d}} \exp \left( -\frac{R}{R_{\rm d}} \right) {\rm sech}^2 \left( \frac{z}{z_{\rm d}} \right),
\end{equation}
with $M_{\rm d}$ the total disc mass, $R_{\rm d}$ and $z_{\rm d}$ the radial and vertical scale lengths. The chosen values for the stellar disc are $M_{\rm star} = 5\times10^{10}\,{\rm M}_{\odot}$, $R_{\rm star} = 3.5$\,kpc and $z_{\rm star} = 0.7$\,kpc. Similarly, the gas disc is characterized by $M_{\rm gas} = 2\times10^{10}\,{\rm M}_{\odot}$, $R_{\rm gas} = 3.5$\,kpc and $z_{\rm gas} = 0.035 z_{\rm star}$. These structural parameters create a late-type disc galaxy broadly comparable in size and mass to the Milky Way.

Mass resolution in the galaxy was set differently for each component based on results from preliminary tests, {in which similar setups were analyzed and verified not to introduce transients or non-physical effects related to resolution, while also maintaining reasonable simulation wall time}. For the dark matter halo, we assigned $N=4\times10^{5}$ particles, each with mass of $2.26\times10^{6}\,{\rm M}_{\odot}$. The bulge was modeled with $N=1.6\times10^{5}$ particles of $1.25\times10^{5}\,{\rm M}_{\odot}$. For the initial stellar disc, we adopted $N=4\times10^{5}$ and $1.24\times10^{5}\,{\rm M}_{\odot}$ for each particle. Finally, due to the nature of {\sc Arepo}'s hydrodynamics, we set a target gas cell mass of $3\times10^{4}\,{\rm M}_{\odot}$ which resulted in $N\approx1.3\times10^6$ gas cells in the galaxy for the following simulations. This target gas cell mass also controlled the entire simulation volume once the galaxy was placed within the final simulation boxes with the gas transitions, and is kept the same across all runs, minimizing transients and non-physical effects between the galaxy and the ICM.

After creating the galaxy it was relaxed for 1\,Gyr in a numerical vacuum with $\rho=10^{-33}$\,g\,cm$^{-3}$, ensuring any measurable transients are dissipated. During the relaxation phase, star formation consumes the gas and the actual $M_{\rm gas}$ inserted for the final simulations was $7.5\times10^{9}\,{\rm M}_{\odot}$ .

\subsection{Galaxy insertion}
\label{subsection:galaxy_insertion}

The galaxy is placed at the center of the simulation volume{, within the base density region $\rho_{\rm{base}}$}, with its disc in the $xy$ plane. Wind is simulated by setting the ICM gas velocity along the $z$ direction to $1000\,\rm{km\,s^{-1}}$ before inserting the galaxy.

{With this setup the galaxy is simulated in its rest frame while the ICM flows past it, transitioning from the base density $\rho_{\rm{base}}$ to the peak density $\rho_{\rm{peak}}$ and back to the initial $\rho_{\rm{base}}$}. However, throughout the text, descriptions will also be given from the ICM rest frame perspective (where the galaxy is moving through the stationary ICM) when convenient. Since a periodic box was used, a maximum of $0.2$\,Gyr is simulated after the initial galaxy relaxation, which is the time the galaxy tail would take to traverse the box length of $200\,{\rm kpc}$ at the ICM velocity of $1000\,\rm{km\,s^{-1}}$. This avoids the effect of the galactic tail interacting with the disc. 

\begin{figure}
    \centering
    \includegraphics{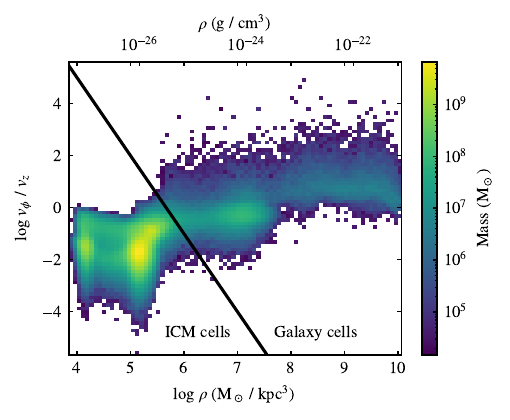}
    \caption{Phase diagram adopted to distinguish galaxy gas cells from the ICM gas. The line has angular coefficient $a=-3$ and linear coefficient $b=19$ for the axes shown, {dividing the phase space into two regions corresponding to ICM and galaxy gas cells, as indicated}. This example is shown for model HD10 at 1.14\,Gyr {(140\,Myr after the initial relaxation)}.}
    \label{fig:ISM_ICM_separation}
\end{figure}

Hydrodynamics in {\sc Arepo} are handled such that gas cells can be merged and refined, while not retaining their history. Because of this, at each snapshot, we identified the gas from the galaxy using a cut in a phase diagram (Fig.~\ref{fig:ISM_ICM_separation}).
In the galaxy's rest frame, its gas is denser and exhibits significant velocity components in the $xy$ plane due to its rotational motion, while the ICM is diffuse and has velocity exclusively in the $z$ direction. {We therefore define the quantity $v_\phi / v_{z}$, which represents the ratio of velocity or momentum in the $xy$ plane relative to the $z$ direction. Along with the density $\rho$, this quantity provides a useful discriminator between galaxy and ICM gas cells.
While verifying different thresholds and methods for distinguishing the regions in Fig.~\ref{fig:ISM_ICM_separation}, we noticed that simple limits in density $\rho$ or the $v_\phi / v_{z}$ ratio (which would correspond to vertical or horizontal lines in Fig.~\ref{fig:ISM_ICM_separation}) were not able to reliably separate gas cells over all models and throughout all snapshots. This can be understood by taking into account that a gas cell from the outer parts of the galaxy can have lower densities comparable to the ICM, but still be dynamically connected to the galactic disc. Compounding thresholds, as in setting some $\psi$ and $\tau$ and defining that the ICM is anything that satisfies $\rho < \psi$ and $v_\phi / v_{z} < \tau$  (effectively creating a box in the lower left region of Fig.~\ref{fig:ISM_ICM_separation}) also presented the same difficulties when setting where exactly these thresholds would be. Empirically, the line depicted in the phase space of Fig.~\ref{fig:ISM_ICM_separation} was verified to work for distinguishing gas cells of the ICM from the galaxy in all simulations and deemed sufficient for our purposes. It is possible that a combination of curves, other lines or a more involved technique such as a clustering algorithm could be applied in this case for similar or better results, but these approaches were considered out of scope.}

\begin{figure}
    \centering
    \includegraphics{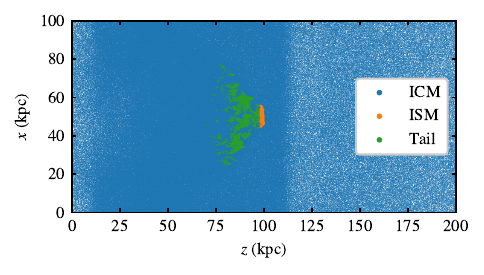}
    \caption{Example of the separation between ICM, ISM and tail gas, shown for model HD10 at 1.14\,Gyr {(140\,Myr after the initial relaxation)}. This scatter shows cell position colored according to the cut in the phase diagram Fig.~\ref{fig:ISM_ICM_separation} and a {distance of $\pm$3\,kpc in the $z$ direction} from the galaxy density peak, defining where the galaxy gas disc ends and the tail starts.}
    \label{fig:separation_fig}
\end{figure}

To distinguish the tail from the remaining galaxy gas disc a simple geometric cut was made considering the tail as being any gas cell outside a distance in $z$ of $3\,{\rm kpc}$ from the galaxy density maximum, as is showcased in Fig.~\ref{fig:separation_fig}. This method provided the means to measure gas mass, SFR and the disc size on the galaxy. Both Figs.~\ref{fig:ISM_ICM_separation} and \ref{fig:separation_fig} correspond to the HD10 simulation at 1.14\,Gyr, chosen arbitrarily to illustrate the method.

\section{Results}
\label{sec:results}

In this section we present the results obtained from the simulations. We start with a qualitative analysis of galaxy morphology, subsequently discussing the measured gas stripping, SFR, and color evolution across all models.

\subsection{Galaxy morphology}

\begin{figure*}
    \centering
    \includegraphics{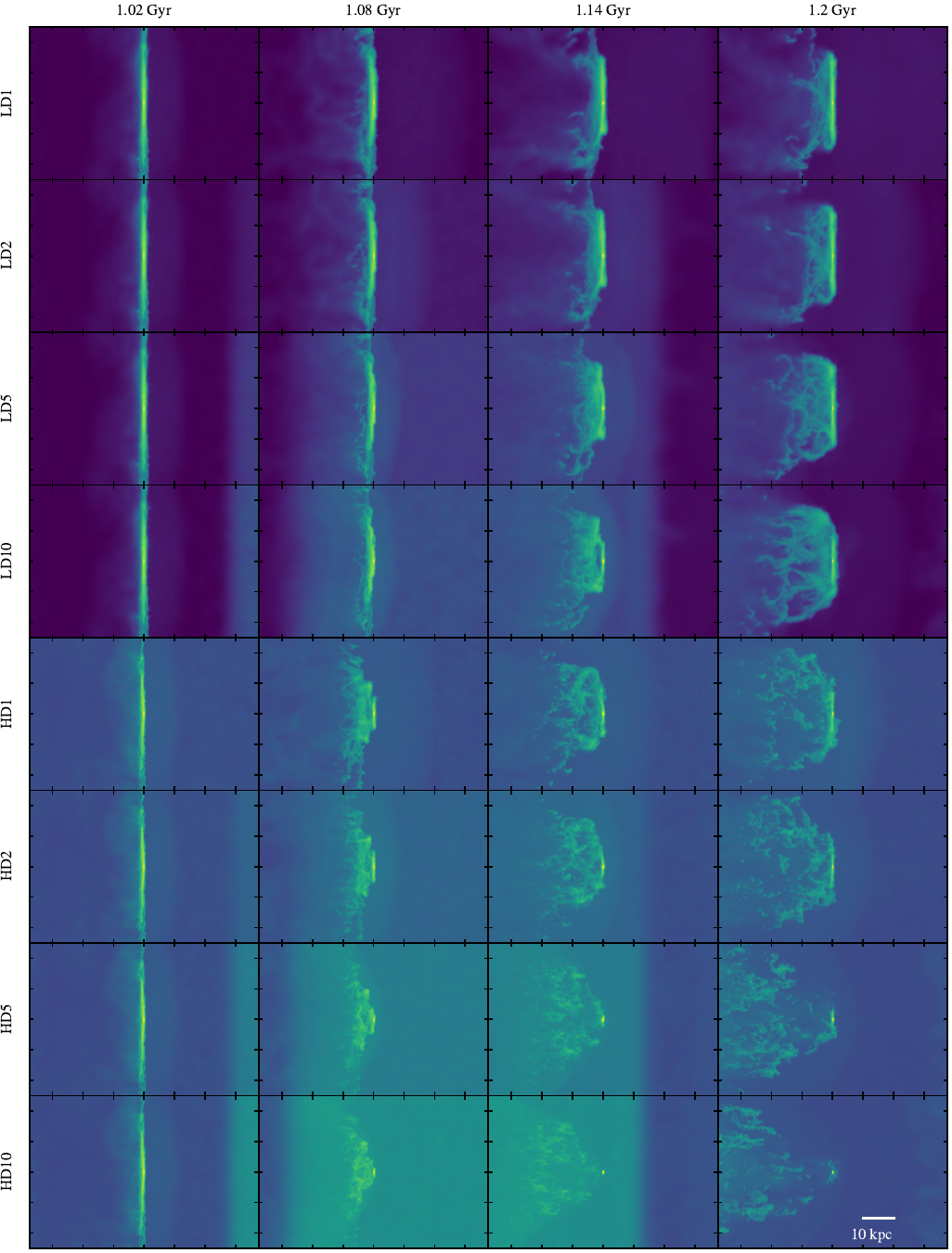}
    \caption{Maps of projected gas density of all simulations. The galaxy disc is seen edge-on, {where each row represents a different model and each column shows equally spaced times.}\\}
    \label{fig:final_simulations}
\end{figure*}


We begin by analyzing the general behavior of the gas over time in all models using maps of projected gas density. In this work we present only edge-on views since the galaxy is always receiving wind face-on and no inclination parameter is explored. In this configuration, jellyfish tails become more prominent, allowing the main differences between models to be more clearly observed.

Fig.~\ref{fig:final_simulations} showcases the gas evolution over time for the 6 models in which the galaxy crosses an interface, with colors representing gas density. {Each column} corresponds to a specific time, equally spaced and set to relevant moments of the simulation: after the relaxation of 1\,Gyr and the first transients of insertion ($t=1.02$\,Gyr), right after crossing the gas interface ($t=1.08$\,Gyr), right before crossing the second interface ($t=1.14$\,Gyr) and at the end of simulation ($t=1.2$\,Gyr). {The rows} correspond to different models in order of the base density and amplitude of transition.

Since ICM velocity is constant, the instants of entry and exit of the galaxy through the denser gas is similar throughout all simulations, with the galaxy disc crossing the center of the first interface at 1.05\,Gyr and the second at 1.15\,Gyr. The ICM setup ensures these transitions are stable in time, which makes their positions identical in all models, except for small numerical fluctuations.

In general, for each column of Fig.~\ref{fig:final_simulations} the galaxy tail constantly grows and the gas disc shrinks with time. Within the same base density --- LD or HD --- the tail becomes more pronounced the stronger the amplitude of transition, going from 1:2 towards 1:10. This trend is also seen across each row, where the tail is increasingly more prominent at a given time going from left to right, showing a subtle tail for LD2 at 1.2\,Gyr and a much more evident tail in the HD10 model. The galaxy disc changes throughout the LD simulations due to ram pressure, but this effect is more visible the stronger the ram pressure experienced, becoming heavily truncated in HD10 as opposed to LD2.


\subsection{Gas stripping}
\label{sec:gas_stripping}
\begin{figure}
    \centering
    \includegraphics{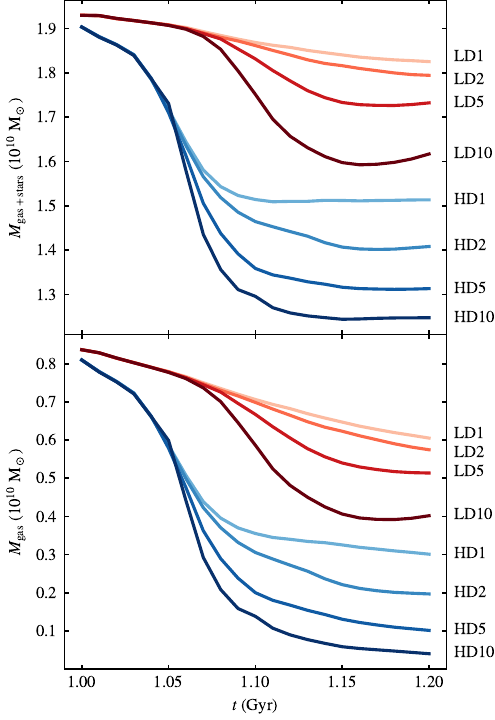}
    \caption{Mass evolution of the {galactic disc} {within $\Delta z \leq$ 3 kpc from the peak density}, considering gas and newly formed star particles {in the upper panel and gas only in the lower panel}. Red tones represent the LD models while blue tones represent HD models. Most mass is removed shortly after 1.05\,Gyr, when the galaxy crosses the transition and enters the densest part of the wind tunnel, which increases ram pressure stripping.}
    \label{fig:mass_evolution}
\end{figure}

One of the most immediate effects of ram pressure on a galaxy is the removal of gas from its disc. This process, known as gas stripping, can alter the star formation activity and long-term evolution of the galaxy. In this section, we present a quantitative analysis of gas stripping across every model by measuring the mass of gas and stars and stripping radius of the galactic disc over time.

Fig.~\ref{fig:mass_evolution} shows the measured mass on the galactic disc for each model, including both formed stellar populations and gas. Red shades represent the LD models, while blue shades correspond to the HD models, with darker colors indicating stronger discontinuities. Accounting for both gas and stellar mass in the disc {isolates} the effects of RPS from apparent mas losses caused by temporary increases in SFR, which could otherwise be misinterpreted when considering only disc gas mass as a function of time. {In the second pannel of Fig.~\ref{fig:mass_evolution} only the disc's gas is shown in function of time, highlighting the continued consumption of gas by the star formation, while the upper pannel remains relatively constant after 1.15\,Gyr}. The disc's gas and star particles were selected using the method described in section \ref{subsection:galaxy_insertion}, considering only those star particles within 3\,kpc from the disc center.

The general behavior for the gas mass evolution on the galaxy show monotonically decreasing mass. All curves are ordered, with LD models at the top portion of Fig.~\ref{fig:mass_evolution} while HD models show more pronounced mass loss and keep to the bottom part of the plot. There is also correlation to the amplitude of transition, where galaxies crossing denser regions lose more mass over time, contributing to the ordered behavior, as should be expected from the stronger ram pressure. {In addition, the measured mass in the first panel of Fig.~\ref{fig:mass_evolution} remains relatively constant for all models once the peak ram pressure ends and models return to a less dense environment where the stripping can't remove any more gas.}

\begin{figure}
    \centering
    \includegraphics{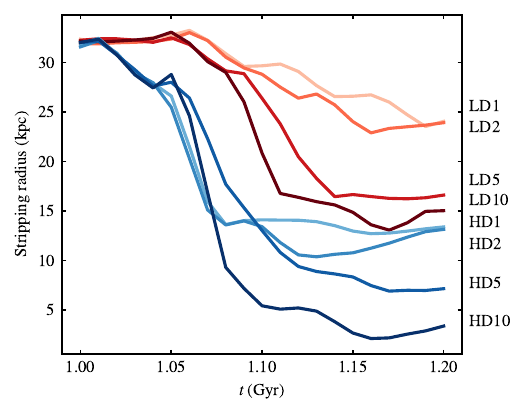}
    \caption{Stripping radius measured on the galactic disc {considering gas within $\Delta z \leq$ 3 kpc from the peak density}. Red shades represent the LD models while blue shades represent HD models. The measured radius decreases after the galaxy in each model crosses the discontinuity, in correlation with the local density it experiences.}
    \label{fig:stripping_radius}
\end{figure}

Fig.~\ref{fig:stripping_radius} shows the measured stripping radius for the galaxy using the method described in \cite{Steinhauser2016}. This approach consists of comparing the actual gas surface density profile of the galaxy at a given moment with the theoretical surface density profile used in the creation of initial conditions. The theoretical surface density profile is recalculated at each snapshot with the same initial scale lengths, but using the remaining gas mass. Stripping radius is then defined as the point where the actual gas surface density drops to one dex below the theoretical profile. This criterion allows for an automated and consistent way quantifying the extent of gas removal from the galactic disc over time.

The initial values of $\sim30\,{\rm kpc}$ in Fig.~\ref{fig:stripping_radius} does not correspond directly to the actual extent of the galactic disc, but rather indicates that the disc had not yet undergone stripping. Likewise, the smaller values observed at later times do not represent the physical extent of the remaining gas disc, but simply reflect that stripping has occurred. {Some fluctuations in the measured radius, particularly near 1.05\,Gyr, are probably produced by changes in the galaxy morphology as a consequence of ram pressure and exacerbated by the transitions, in the case of XD2-10, and possible initial transients}. Nonetheless, this metric provides a basis for comparing the different simulations.

{All stripping radius curves in} Fig.~\ref{fig:stripping_radius} behave similarly to the mass evolution in Fig.~\ref{fig:mass_evolution}. In general, they feature an accelerated decrease in radius when the galaxy interacts with the densest part of each model from 1.05\,Gyr to 1.15\,Gyr, followed by a phase of relative stability, also being ordered according to the density of their respective environments.

{As is shown in Fig.~\ref{fig:mass_evolution} and \ref{fig:stripping_radius} all models have similar initial values of mass and stripping radius at the beginning of the tunnel, but quickly diverge into two groups after this initial moment due to the different environments in LD and HD simulations. Although the effect of the transition is more quickly seen in the HD models at 1.05\,Gyr it also affects LD models and further separates each individual simulation immediately after. This is the primary difference in lost gas mass between the base models --- LD1 and HD1 --- and their counterparts with density transitions, where the curves begin to diverge as each galaxy interacts with the densest region of the wind tunnel after 1.05\,Gyr}.

{Having a closer look at the curves in both} Figs.~\ref{fig:mass_evolution} and \ref{fig:stripping_radius}, it can be seen that the LD1 galaxy experiences a relatively linear decline in mass and radius throughout the simulation, due to its constant gas distribution. Although HD1 shares a proportional gas profile, the denser environment leads to a sharper decrease in mass and stripping radius over less than 0.1\,Gyr. 



After crossing the denser region, LD5 and LD10 galaxies exhibit a small increase in mass of 1$\times10^{8}\,{\rm M}_\odot$ and 2$\times10^{8}\,{\rm M}_\odot$, respectively. This amounts to $\sim1\%$ increase in the combined mass of gas and stars in the galactic disc for both cases, and may be related to gas accretion occurring after the ram pressure has subsided. However, this effect will not be considered for our analysis, as we focus on the more significant phenomena more directly correlated with the crossing of a sloshing spiral.

\subsection{Star formation rate}

In addition to gas loss, interacting with the ICM can affect the star formation activity in galaxies. We now investigate how the variations in density in our tunnel may affect the SFR evolution in each simulation.

\begin{figure}
    \centering
    \includegraphics{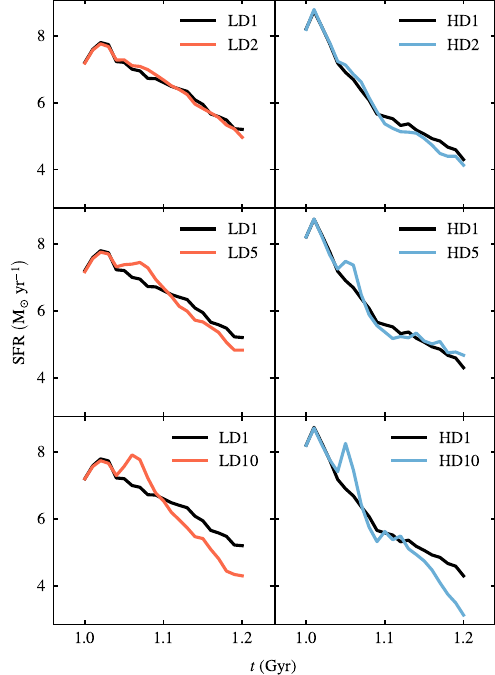}
    \caption{Star formation rate measured in the galactic disc. Left panels display the LD simulations while the right hand side plots correspond to HD. Black curves represent the control simulations LD1 and HD1, while the colored curves represent the models with discontinuities. The models are ordered according to the transition amplitude, going from the LD2 and HD2 models at the upper plots to LD10 and HD10 at the bottom plots.}
    \label{fig:sfr_evolution}
\end{figure}


The SFR measured on the galactic disc is shown in Fig.~\ref{fig:sfr_evolution}. The left-hand side panels display LD models, while right-hand side panels contain the HD counterpart. For this analysis, gas cells were selected in the same way as in section \ref{sec:gas_stripping}. For each plot, the black curve represents the SFR measured in the base models (LD1 and HD1) with a constant density wind tunnel, while the colored curves represent the models with discontinuities. The plots are arranged according to the amplitude of transition, with mildest cases, LD2 and HD2, shown at the top and the strongest cases, LD10 and HD10, at the bottom panels.

After the brief initial transient when the galaxy crosses the first wind tunnel interface at 1.05\,Gyr there is an increase in SFR over all simulations in comparison to the base models. This enhancement is very mild in the LD2 and HD2 galaxies, but becomes appreciable for the more extreme models. The SFR then decreases faster, becoming lower than the base SFR in less than 0.1\,Gyr. For LD models, this effect seems to last even after the galaxy exits the denser region, whereas {HD2 \& 5} galaxies regain approximately similar values to the base model in the same time frame.


HD10 simulation exhibits a different behavior compared to the other models after exiting the denser section, with a significantly sharper drop in SFR. This can be attributed to the rapid stripping and truncation of its gas disc, as displayed in Figs.~\ref{fig:final_simulations}, \ref{fig:mass_evolution} and \ref{fig:stripping_radius}, {which greatly lessens the available cold gas}. Quantitatively, {only $\sim$5\%} of the initial gas mass is left on the disc for {HD10, almost completely starving the galaxy from star forming gas,} {while} {the HD1 simulation retains $\sim$37\%} of its initial gas mass at the end of the run in comparison.


\subsection{Color evolution}

The color index of a galaxy encodes valuable information for a wide variety of properties of a stellar population, including their age, star formation history and metallicity. As such, it can be a useful probe for assessing the effects a galaxy undergoes when interacting with discontinuities in the ICM, also providing measurable values that can be compared with real observations. In this section we give a brief explanation of the method involved for emulating the color index of the galaxy and analyze the behavior of this quantity in each model.

In order to compute synthetic color indices from the simulations, we employed the Flexible Stellar Population Synthesis (FSPS) code \citep{Conroy2010}. We configured the code using the default libraries, selecting its Padova database for isochrones due to the better resolution in metallicity. FSPS makes it possible to compute the spectra and magnitudes of stellar populations given certain inputs about its properties. In our specific case, we set a look up table for mapping the metallicity and age of each star particle generated during the simulation, considered here as a simple stellar population. The output is expressed as magnitudes in the Sloan Digital Sky Survey (SDSS) filters. In particular, we focus on the evolution of the $u-i$ color index.

\begin{figure}
    \centering
    \includegraphics{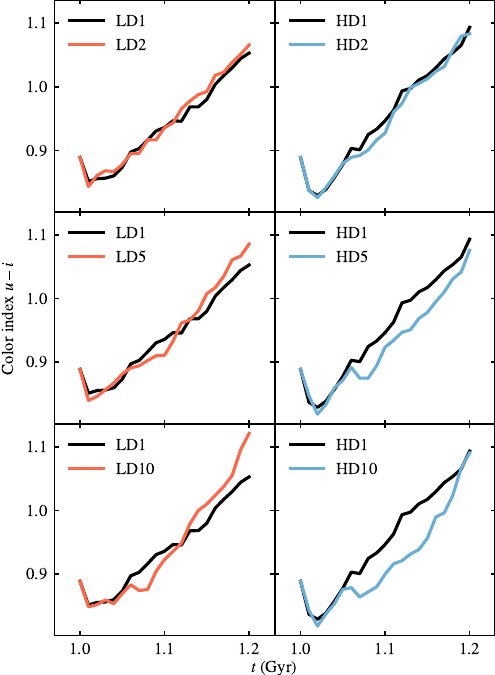}
    \caption{Galaxy integrated color index evolution measured on the stellar disc. Left panels display the LD simulations while the right hand side plots correspond to HD. Black curves represent the control simulations LD1 and HD1, while the colored curves represent the models with discontinuities. The models are ordered according to the transition amplitude, going from the LD2 and HD2 models at the upper plots to LD10 and HD10 at the bottom plots.}
    \label{fig:color_evolution}
\end{figure}

Fig.~\ref{fig:color_evolution} shows the integrated color index measured from the stellar components using the {\sc FSPS} code. The panels are organized in the same way as Fig.~\ref{fig:sfr_evolution}, showing LD models on the left and HD models on the right, with increasing transition amplitude from top to bottom. We assumed that the pre-existing particles from the bulge and stellar disc had an age of 7\,Gyr with metallicity of 1\,Z$_\odot$, while considering the simulated metallicities and formation times for the populations created during simulation. These assumptions are arbitrary, but reflect some of the broader properties of Milky Way-like galaxies and provide a consistent baseline for comparison between models. Moreover, the focus lies on the relative comparison between models rather than absolute values. The analysis remains robust, since the assumptions of ages and metallicities for the older stellar population are the same in all models.

After the initial transient, simulations undergo a quick decrease in color index after 1.05\,Gyr of total simulation time, which coincides with the crossing of the first ICM interface. This is evident for all models except for LD2, which remains mostly undistinguishable from LD1. Aside from the LD10 and HD10 curves, all values for color index remain within 5\% of each other for the rest of simulation time, quickly becoming similar {after the mark of} 1.10\,Gyr.

The most notable differences in color happen for more extreme models, such as LD10, HD5 and HD10. Specifically, the LD10 galaxy has its color index becoming significantly redder at the end of simulation than its counterpart that crosses a constant density gas. For HD5 and HD10, their color is appreciably bluer while the simulation lasts.



\section{Discussion}
\label{sec:discussion}


Our idealized setup provided the means to probe what kind of changes can be expected from a galaxy that crosses an environment comparable to a sloshing spiral. As suggested by Fig.~\ref{fig:sloshing}, realistic self-consistent density and temperature profiles may have less abrupt transitions than our idealized profiles seen in Fig.~\ref{fig:box_profiles}. However, we eliminate any external influences while isolating the process of a galaxy abruptly interacting with colder and denser gas for a brief time. While not extremely realistic, it still serves as a controlled study that can provide insight to the {complete} phenomenon. We discuss and relate our results with previous work in Section \ref{subsection:sumary_comparison} and {specific setup aspects and computational restraints are discussed in Section \ref{subsecion:limitations} respectively}.


\subsection{Impact of cold fronts and comparison with previous work}
\label{subsection:sumary_comparison}

Results show that gas stripping and SFR are influenced by the transition to a denser gas. For a brief time after the abrupt environment change the SFR is significantly enhanced, which subsequently impacts the galaxy's color index. In more extreme cases, such as LD10 and HD10 models, these effects can last longer and appreciably change the color index with relation to a galaxy that traverses a regular ICM with no discontinuities. Nevertheless, less extreme models hardly show changes in color index and SFR that surpass 5\% difference from the base model, and at the final simulation time most curves display similar values to each other. This hints that although the effect of a sloshing spiral can be perceived on a galaxy if extreme enough, it is a short lived and difficult to detect phenomenon.

Astrophysical hydrodynamical simulations have not yet come to a clear consensus whether the increasing ram pressure and ambient density enhances SFR in the remaining gas disc. This has been discussed extensively with contradicting results and different methods \citep{Kapferer2008, Kapferer2009, Kronberger2008, Roediger2014, Steinhauser2016, Lee2020}. In our models, we observe a consistent initial increase in SFR that lasts for less than 0.1\,Gyr when the galaxy transitions to the denser environment in comparison with the control run {in Fig.~\ref{fig:sfr_evolution}}. This is followed by a decline in SFR across all LD simulations. In contrast, after the initial increase, no significant decrease is observed in relation to the control run in the HD2 and HD5 galaxies. Considering the color index these effects result in the reddening of LD models, while HD models {remain} bluer for most of the simulation time and seem to be increasing their index towards redder colors by the end of each run.

On the other hand, the mass evolution of the galactic disc shows the most apparent differences between all models, as discussed in Section~\ref{sec:gas_stripping}. It is strongly related to the amplitude of the discontinuity as a consequence of the ram pressure a galaxy experiences. It is also tied to the stripping radius from Fig.~\ref{fig:stripping_radius}, where the stronger amplitude of transition implies in more mass lost on the disc and a smaller stripping radius.

Even though it was not analyzed in detail, the peculiar increase in mass measured in Section~\ref{sec:gas_stripping} after the galaxy exits the denser region of the simulation may be related to a redistribution of the ISM. A similar effect is described in \cite{Steinhauser2016}, in which one of the RPS models shows its gas disc being redistributed as soon as the ram pressure weakens, a case comparable to ours. \cite{Tonnesen2009} also found that non-stripped gas may be compressed into higher density clouds and drift towards the galactic center as it loses angular momentum in low ram pressure environments. Such a redistribution could cause gas cells around the galaxy to change their properties and traverse the phase diagram of Section~\ref{subsection:galaxy_insertion}, leading them to be mistakenly considered as ISM. Another possible explanation to the measured increase in mass is simply the gas accretion into regions that can still bound gas gravitationally but were being stripped while in the denser regions.

Analyzing LD5 and LD10 disc mass, SFR and color index graphs (Figs.~\ref{fig:mass_evolution}, \ref{fig:sfr_evolution} and \ref{fig:color_evolution}), SFR increases and the color index becomes bluer initially {due to the recent star formation}. However, this effect then reverses around the 1.15\,Gyr mark when transitioning to a less dense tunnel region, where the galaxies then become increasingly redder than their counterpart crossing constant density. These models preserve a considerable part of the gas disc throughout the simulation, suggesting that if the ram pressure stripping isn't strong enough to quickly destroy the disc --- as seen mainly in the HD10 model with Figs.~\ref{fig:final_simulations} and \ref{fig:stripping_radius}--- observable galaxies that went through a sloshing spiral may have a redder color index, less SFR and gas content than a similar galaxy in an undisturbed medium. {HD5 and HD10 models show a similar initial bluer index that increases towards redder values in later simulation stages. It is possible that the initial higher SFR due to the more dense medium and the SFR peak at 1.05\,Gyr produces enough stellar populations for keeping the bluer color for longer. This hints that the same process operating in LD5 and LD10 is also acting here and that HD5 and HD10 systems would eventually become redder given sufficient evolution time, as their stellar populations age}.

In \cite{Goller2023} a set of jellyfish galaxies is analyzed using IllustrisTNG data \citep{Nelson2019} and their results show that even though bursts of SFR happen in these galaxies, changes are not population-wide. Our findings are consistent with this picture, in the sense that our simulations also predict that there should not be a distinct population of exceptionally star-forming jellyfish galaxies, even in merging clusters, {since SFR was measured to very briefly increase, but only when crossing more intense cold fronts, which should occur for only a fraction of galaxies in a cluster.}

{Also concerning SFR, although very deprived of star forming gas, the HD10 galaxy keeps producing new stars at a reduced rate until the end of the run, having lost $95\%$ of its original gas content, which is compatible with \cite{Rohr2023} findings that jellyfish galaxies can keep forming stars even when $\sim 98\%$ of their cold gas mass is removed.}

In the HD5 and HD10 models, with the quick SFR spike, the color indices become bluer and retains a lower value until the simulation end. Due to the much stronger ram pressure stripping, these galaxies have their discs quickly truncated. This suggests that a galaxy crossing more extreme sloshing spirals or denser regions of a cluster will become bluer, at least for a brief time after crossing it, having its gas disc heavily truncated and showing even more discernible tails of gas. 

In our milder simulations LD2 and HD2, hardly any difference can be noted in comparison to the base models. There are some fluctuations in SFR and color, more mass from the disc is also lost and the stripping radius is smaller than the base model while crossing the denser simulation region. Despite that, all changes to the galaxy are not as significant as those seen in the other models. Considering LD2 and HD2 initial conditions setup, their evolution imply that a sloshing spiral or shock discontinuity should be more than twice as dense in its center than its surrounding to significantly change the evolution of a galaxy crossing it.

{Another comparison worth considering is how similar the evolution of galaxies LD10 and HD1 is, since they both experience the same peak ram pressure. This raises the possibility that the effects of a cold front passage are degenerate with those of a pericentric passage.} Even though in both cases the same peak ram pressure influences the galaxy, the shorter period it acts on LD10 causes it to lose less gas (Fig.~\ref{fig:mass_evolution}). Given enough time within the denser region, it is reasonable to expect that LD10 would lose additional gas and plateau around the same $M_{\rm gas + stars}$ value, behaving similarly to HD1, {especially} considering the general behavior of all models. However, LD simulations are meant to represent a less dense and more external region of a cluster, implying that this aspect in fact shows that cold fronts could exacerbate the ram pressure stripping in galaxies not close to the center of a cluster, causing comparable effects to a pericentric passage (see Section~\ref{sec:comparison_cluster}). This suggests that cold fronts may partially reproduce the stripping conditions of denser cluster environments without requiring the galaxy to reach the innermost regions of the cluster. Therefore, not only the peak ram pressure, but also the {duration of the} interaction appear to be relevant.



These findings correlate to \cite{Ruggiero2019}, where it is proposed that more extreme jellyfish galaxies are preferentially located near boundaries in cluster mergers. These regions would contain enhanced properties that could considerably change ram pressure, such as the density {transitions} imposed in our models.

\cite{Lourenco2023} finds a slight increase in the fraction of galaxies undergoing RPS in interacting clusters. Although many caveats are drawn in their description, such as the difficulties of observing interacting clusters, the small sample size, and the inherent complexity of mergers, it was believed that RPS features would be more common in these disturbed systems. Given our results, it can be argued that galaxy populations are not expected to be strongly affected in merging clusters due to the short-lived aspect seen in all phenomena analyzed, where most changes happen in fractions of a Gyr. Nonetheless, the exact merger configuration may also have an important role, where only specially extreme cases of shocks or cold fronts comparable to the HD10 and LD10 models are able to significantly change the fraction of stripped galaxies. 

Notwithstanding, real galaxies traversing such regions can be difficult to observe and classify due to the merger geometry \citep{Poggianti2016, Lourenco2023}, which also poses a difficult scenario for real comparisons and analysis. {Sloshing spirals are not rare. In fact, \cite{Lagana2010} estimated that nearly half of the clusters from a nearby sample showed signs of spiral-like structures. However, orientation plays a key role. The density excess of a sloshing spiral is a three-dimensional feature, but it is in fact more intense in the plane of the orbit of the perturber that induced the sloshing. In other words, from the observational point of view, the sloshing phenomenon is more clearly seen when the orbital plane of the perturbation is close to the plane of the sky. Thus, our simulated galaxies would need to be moving mostly in the plane of the sky. And since they enter the cold front face-on, this means that one would observe them is the edge-on projection. This would be the most favorable configuration to allow detection of both the cluster cold front, and the jellyfish tails. If, on the other hand, the merger is highly inclined away from the plane of the sky, the cold front projected morphology becomes complicated \citep{Machado2022} and the detection of jellyfish tails might also be hindered. These are factors contributing to the decreased likelihood of detection of this phenomenon.}

\subsection{Comparison with typical cluster conditions}
\label{sec:comparison_cluster}

{
To place our findings in a more observational context, we consider the local ram pressure experienced by a galaxy for a certain density $\rho$ and relative velocity $v$ in the ICM \citep{Gunn1972},
\begin{equation}
    P_{\rm ram} = \rho v^2.
\end{equation}
We take into account the Coma cluster (Abell 1656), one of the most well known clusters in our Universe \citep{Biviano1998}. The Coma cluster has a virial mass of around $10^{15}\,{\rm M_{\odot}}$ \citep{Briel1992}, which roughly corresponds to our motivation model in Section\,\ref{sec:sloshing_spirals} and has a velocity dispersion of $\sigma=1082\,{\rm km\,s^{-1}}$ \citep{Colless1996}. Its ICM density distribution can be described through the $\beta$ model proposed by \cite{Briel1992}:
\begin{equation}
\label{eq:beta_model}
    \rho(r) = \rho_0 \times \left[ 1 + \left(\frac{r}{r_c}\right)^2 \right]^{-3\beta/2}
\end{equation}
with $\beta = 3/4$, $r_c = 420\,{\rm kpc}$ the core radius and $\rho_0 = 5.7\times10^{-27}\,{\rm g\,cm^{-3}}$ the scale density \citep{Briel1992, Neumann2003}.
With this parameterization and adopting a representative velocity of $v=1000\,{\rm km\,s^{-1}}$--- similar to both the cluster dispersion and that of our simulations, a galaxy during a pericentric passage at $r=300\,{\rm kpc}$ in the Coma cluster would experience
\begin{equation}
    P_{\rm ram} \approx 3.6\times10^{-11}{\rm\,dyn\,cm^{-2}}.
\end{equation}
This value is comparable to the ram pressure reached in our HD models, where peak values range from
\begin{equation}
    P_{\rm ram} \approx 10^{-11} ~\text{---}~ 10^{-10}{\rm\,dyn\,cm^{-2}}.
\end{equation}
This suggests that our strongest models can reach and even exceed the ram pressure expected during a typical pericentric passage in this scenario. It also contributes to the idea that crossing cold fronts that are sufficiently denser than the surrounding environment may be capable of significantly disrupting the evolution of a jellyfish galaxy --- at least temporarily, until it further interacts with other galaxies or observations become indistinguishable from ordinary cluster processing, as was suggested when comparing LD10 and HD1 previously.
}

\subsection{Limitations of our simulations}
\label{subsecion:limitations}

Although the simulation setup strove for accurate and controlled conditions, some limitations arise from the employed methodology and numerical constraints. A transient effect can be seen at the beginning of all curves in Figs.~\ref{fig:sfr_evolution} and \ref{fig:color_evolution} for about $0.02\,{\rm Gyr}$. {Even though this transient seems to be more extreme for HD models, all} simulations are able to sufficiently relax before crossing the discontinuities at 1.05\,Gyr, where the effects from entering the denser region are much more significant. At the transient, the star formation model produces a negligible amount of stellar mass and results in equally small disturbances in color index, but any later important interaction completely outweighs the effects of this brief moment. Furthermore, the comparison of each run with the base model isolates the important features of later simulations times, lessening any effect these transients could have in our analysis.

We employed the specific simulation volume of $x \times y \times z = 100\,{\rm kpc} \times 100\,{\rm kpc} \times 200\,{\rm kpc}$ to obtain the maximum mass resolution possible while running all eight models in a reasonable amount of time with our available computational resources. This ultimately limited the run time for the galaxy interacting with the ICM wind to 0.2\,Gyr, which is the time the galaxy takes to loop around the periodic boundaries and encounter its tail. Nevertheless, a larger simulation volume could stray from the case where a galaxy interacts with a cluster: as is concluded in \cite{Tonnesen2019}, a constant wind cannot properly model this case, since the specific trajectory and local density of the ICM changes the galaxy evolution. {The adoption of constant density and velocity is a meaningful limitation of the current work. Therefore, the simulations presented here are meant to represent a relatively short segment of an orbit, along which the assumption of constant density might be considered an acceptable first approximation.} We can compare our case to a galaxy crossing a small enough section of a cluster where properties are similar to the gas profiles we applied to our simulation volume. {This also isolates the cold front structure from other phenomena present in the complex interactions of galaxies and galaxy cluster collisions, reducing our parameter space and allowing for a more focused outlook.}

Another important aspect is that cold fronts evolve and may change their morphology over time, as well as their density and temperature profiles. Our simulations already account for the relative velocity a galaxy may have with a sloshing spiral by using the wind tunnel setup. Even so, all models interact with a non time evolving cold front, by construction. This is justified by noting that cold fronts persist and evolve for multiple Gyr \citep{Ascasibar2006}, whereas our idealized case is simulated for only 0.2\,Gyr. Over this time span, it is an acceptable approximation to consider cold front static.


\section{Conclusions}
\label{sec:conclusions}

In this paper we analyzed a series of simulations in order to identify possible changes a sloshing spiral may impose onto a galaxy. These simulations were composed of an idealized galaxy simulated with a wind tunnel setup that recreated the discontinuities in temperature and density seen in sloshing cold fronts with an analytical gas profile along the tunnel.

Our analysis revealed that changes in color and SFR are typically subtle. Most models barely reached 5\% difference in these quantities when compared to control simulations. In such cases, most induced phenomena were also short-lived. On the more extreme models, the galactic disc was quickly truncated due to the denser medium after the first transition, showing alterations that persisted through the entire simulation run, even though color index and SFR eventually return to values comparable to those of the control runs.

Since the effects are mild and short-lived, these simulations suggest that, when comparing sloshing clusters and relaxed clusters, their populations of jellyfish galaxies are not expected to be significantly different --- at least not as a consequence of ram pressure within weaker cold fronts. However, clusters with stronger cold fronts --- where gas density increases by factors of at least 5-10 times across the interface --- may host galaxies with exceptionally truncated gas discs.




The wind tunnel setup with an idealized profile for the gas properties developed in this work could be used in future investigations of jellyfish galaxies interacting with ICM discontinuities. An expansion of the parameter space adding different galaxy inclinations, ICM velocities and different structural parameters of the galaxy itself, such as stellar mass, gas content and morphology would provide a better understanding of this phenomenon. Longer simulation times in more realistic environments could provide more insight into long-lasting effects. Additionally, mock images of X-ray and H$\alpha$ would supply more direct comparisons with observable galaxies.

\section*{Acknowledgements}

The authors thank V. Springel for making available the non-public version of the {\sc Arepo} code and L. P. Martins for insights about stellar population synthesis. EAMT acknowledges support from UTFPR. This study was financed in part by the Coordenação de Aperfeiçoamento de Pessoal de Nível Superior - Brasil (CAPES) - Finance Code 001. REGM acknowledges support from \textit{Conselho Nacional de Desenvolvimento Cient\'ifico e Tecnol\'ogico} (CNPq) through grants 406908/2018-4 and 303425/2024-5, and from \textit{Funda\c c\~ao de Apoio \`a Ci\^encia, Tecnologia e Inova\c c\~ao do Paran\'a} through grant 18.148.096-3 --- NAPI \textit{Fen\^omenos Extremos do Universo}. RPA acknowledges financial support from FAPESP under grant 2024/13224-9. 


\bibliographystyle{aasjournal}
\bibliography{bibliography}

@ARTICLE{Ascasibar2006,
       author = {{Ascasibar}, Yago and {Markevitch}, Maxim},
        title = "{The Origin of Cold Fronts in the Cores of Relaxed Galaxy Clusters}",
      journal = {\apj},
         year = 2006,
        month = oct,
       volume = {650},
       number = {1},
        pages = {102-127},
          doi = {10.1086/506508},
archivePrefix = {arXiv},
       eprint = {astro-ph/0603246},
 primaryClass = {astro-ph},
       adsurl = {https://ui.adsabs.harvard.edu/abs/2006ApJ...650..102A}
}

@ARTICLE{Larson1980,
       author = {{Larson}, R.~B. and {Tinsley}, B.~M. and {Caldwell}, C.~N.},
        title = "{The evolution of disk galaxies and the origin of S0 galaxies}",
      journal = {\apj},
         year = 1980,
        month = may,
       volume = {237},
        pages = {692-707},
          doi = {10.1086/157917},
       adsurl = {https://ui.adsabs.harvard.edu/abs/1980ApJ...237..692L}
}

@ARTICLE{Balogh2000,
       author = {{Balogh}, Michael L. and {Navarro}, Julio F. and {Morris}, Simon L.},
        title = "{The Origin of Star Formation Gradients in Rich Galaxy Clusters}",
      journal = {\apj},
         year = 2000,
        month = sep,
       volume = {540},
       number = {1},
        pages = {113-121},
          doi = {10.1086/309323},
archivePrefix = {arXiv},
       eprint = {astro-ph/0004078},
 primaryClass = {astro-ph},
       adsurl = {https://ui.adsabs.harvard.edu/abs/2000ApJ...540..113B}
}

@ARTICLE{Machado2022,
       author = {{Machado}, R.~E.~G. and {Lagan{\'a}}, T.~F. and {Souza}, G.~S. and {Caproni}, A. and {Antas}, A.~S.~R. and {Mello-Terencio}, E.~A.},
        title = "{Simulating nearly edge-on sloshing in the galaxy cluster Abell 2199}",
      journal = {\mnras},
         year = 2022,
        month = sep,
       volume = {515},
       number = {1},
        pages = {581-593},
          doi = {10.1093/mnras/stac1829},
archivePrefix = {arXiv},
       eprint = {2206.14127},
 primaryClass = {astro-ph.GA},
       adsurl = {https://ui.adsabs.harvard.edu/abs/2022MNRAS.515..581M}
}

@ARTICLE{Ruggiero2017,
   author = {{Ruggiero}, R. and {Lima Neto}, G.~B.},
    title = "{The fate of the gaseous discs of galaxies that fall into clusters}",
  journal = {\mnras},
archivePrefix = "arXiv",
   eprint = {1703.08550},
     year = 2017,
    month = jul,
   volume = 468,
    pages = {4107-4115},
      doi = {10.1093/mnras/stx744},
   adsurl = {http://adsabs.harvard.edu/abs/2017MNRAS.468.4107R}
}

@ARTICLE{Ruggiero2019,
       author = {{Ruggiero}, Rafael and {Machado}, Rubens E.~G. and {Roman-Oliveira}, Fernanda V. and {Chies-Santos}, Ana L. and {Lima Neto}, Gast{\~a}o B. and {Doubrawa}, Lia and {Rodr{\'\i}guez del Pino}, Bruno},
        title = "{Galaxy cluster mergers as triggers for the formation of jellyfish galaxies: case study of the A901/2 system}",
      journal = {\mnras},
         year = 2019,
        month = mar,
       volume = {484},
       number = {1},
        pages = {906-914},
          doi = {10.1093/mnras/sty3422},
archivePrefix = {arXiv},
       eprint = {1812.05611},
 primaryClass = {astro-ph.GA},
       adsurl = {https://ui.adsabs.harvard.edu/abs/2019MNRAS.484..906R}
}

@ARTICLE{Springel2003,
       author = {{Springel}, Volker and {Hernquist}, Lars},
        title = "{Cosmological smoothed particle hydrodynamics simulations: a hybrid multiphase model for star formation}",
      journal = {\mnras},
         year = 2003,
        month = feb,
       volume = {339},
       number = {2},
        pages = {289-311},
          doi = {10.1046/j.1365-8711.2003.06206.x},
archivePrefix = {arXiv},
       eprint = {astro-ph/0206393},
 primaryClass = {astro-ph},
       adsurl = {https://ui.adsabs.harvard.edu/abs/2003MNRAS.339..289S}
}

@ARTICLE{Hernquist1990,
   author = {{Hernquist}, L.},
    title = "{An analytical model for spherical galaxies and bulges}",
  journal = {\apj},
     year = 1990,
    month = jun,
   volume = 356,
    pages = {359-364},
      doi = {10.1086/168845},
   adsurl = {http://adsabs.harvard.edu/abs/1990ApJ...356..359H}
}

@ARTICLE{Lagana2010,
   author = {{Lagan{\'a}}, T.~F. and {Andrade-Santos}, F. and {Lima Neto}, G.~B.
	},
    title = "{Spiral-like structure at the centre of nearby clusters of galaxies}",
  journal = {\aap},
archivePrefix = "arXiv",
   eprint = {0911.3785},
 primaryClass = "astro-ph.CO",
     year = 2010,
    month = feb,
   volume = 511,
      eid = {A15},
    pages = {A15},
      doi = {10.1051/0004-6361/200913180},
   adsurl = {http://adsabs.harvard.edu/abs/2010A%26A...511A..15L}
}

@ARTICLE{Akerman2023,
       author = {{Akerman}, Nina and {Tonnesen}, Stephanie and {Poggianti}, Bianca Maria and {Smith}, Rory and {Marasco}, Antonino},
        title = "{How Ram Pressure Drives Radial Gas Motions in the Surviving Disk}",
      journal = {\apj},
         year = 2023,
        month = may,
       volume = {948},
       number = {1},
          eid = {18},
        pages = {18},
          doi = {10.3847/1538-4357/acbf4d},
archivePrefix = {arXiv},
       eprint = {2301.09652},
 primaryClass = {astro-ph.GA},
       adsurl = {https://ui.adsabs.harvard.edu/abs/2023ApJ...948...18A}
}

@ARTICLE{Gunn1972,
       author = {{Gunn}, James E. and {Gott}, J. Richard, III},
        title = "{On the Infall of Matter Into Clusters of Galaxies and Some Effects on Their Evolution}",
      journal = {\apj},
         year = 1972,
        month = aug,
       volume = {176},
        pages = {1},
          doi = {10.1086/151605},
       adsurl = {https://ui.adsabs.harvard.edu/abs/1972ApJ...176....1G}
}

@ARTICLE{Kronberger2008,
       author = {{Kronberger}, T. and {Kapferer}, W. and {Ferrari}, C. and {Unterguggenberger}, S. and {Schindler}, S.},
        title = "{On the influence of ram-pressure stripping on the star formation of simulated spiral galaxies}",
      journal = {\aap},
         year = 2008,
        month = apr,
       volume = {481},
       number = {2},
        pages = {337-343},
          doi = {10.1051/0004-6361:20078904},
archivePrefix = {arXiv},
       eprint = {0801.3759},
 primaryClass = {astro-ph},
       adsurl = {https://ui.adsabs.harvard.edu/abs/2008A&A...481..337K}
}

@ARTICLE{Roediger2014,
       author = {{Roediger}, E. and {Bruggen}, M. and {Owers}, M.~S. and {Ebeling}, H. and {Sun}, M.},
        title = "{Star formation in shocked cluster spirals and their tails.}",
      journal = {\mnras},
         year = 2014,
        month = sep,
       volume = {443},
        pages = {L114-L118},
          doi = {10.1093/mnrasl/slu087},
archivePrefix = {arXiv},
       eprint = {1405.1033},
 primaryClass = {astro-ph.GA},
       adsurl = {https://ui.adsabs.harvard.edu/abs/2014MNRAS.443L.114R}
}

@ARTICLE{Roediger2006,
       author = {{Roediger}, Elke and {Br{\"u}ggen}, Marcus},
        title = "{Ram pressure stripping of disc galaxies: the role of the inclination angle}",
      journal = {\mnras},
         year = 2006,
        month = jun,
       volume = {369},
       number = {2},
        pages = {567-580},
          doi = {10.1111/j.1365-2966.2006.10335.x},
archivePrefix = {arXiv},
       eprint = {astro-ph/0512365},
 primaryClass = {astro-ph},
       adsurl = {https://ui.adsabs.harvard.edu/abs/2006MNRAS.369..567R}
}

@ARTICLE{Roediger2005,
       author = {{Roediger}, E. and {Hensler}, G.},
        title = "{Ram pressure stripping of disk galaxies. From high to low density environments}",
      journal = {\aap},
         year = 2005,
        month = apr,
       volume = {433},
       number = {3},
        pages = {875-895},
          doi = {10.1051/0004-6361:20042131},
       adsurl = {https://ui.adsabs.harvard.edu/abs/2005A&A...433..875R}
}

@ARTICLE{Steinhauser2012,
       author = {{Steinhauser}, D. and {Haider}, M. and {Kapferer}, W. and {Schindler}, S.},
        title = "{Galaxies undergoing ram-pressure stripping: the influence of the bulge on morphology and star formation rate}",
      journal = {\aap},
         year = 2012,
        month = aug,
       volume = {544},
          eid = {A54},
        pages = {A54},
          doi = {10.1051/0004-6361/201118311},
archivePrefix = {arXiv},
       eprint = {1208.1265},
 primaryClass = {astro-ph.CO},
       adsurl = {https://ui.adsabs.harvard.edu/abs/2012A&A...544A..54S}
}

@ARTICLE{Steinhauser2016,
       author = {{Steinhauser}, Dominik and {Schindler}, Sabine and {Springel}, Volker},
        title = "{Simulations of ram-pressure stripping in galaxy-cluster interactions}",
      journal = {\aap},
         year = 2016,
        month = jun,
       volume = {591},
          eid = {A51},
        pages = {A51},
          doi = {10.1051/0004-6361/201527705},
archivePrefix = {arXiv},
       eprint = {1604.05193},
 primaryClass = {astro-ph.GA},
       adsurl = {https://ui.adsabs.harvard.edu/abs/2016A&A...591A..51S}
}

@ARTICLE{Tonnesen2007,
       author = {{Tonnesen}, Stephanie and {Bryan}, Greg L. and {van Gorkom}, J.~H.},
        title = "{Environmentally Driven Evolution of Simulated Cluster Galaxies}",
      journal = {\apj},
         year = 2007,
        month = dec,
       volume = {671},
       number = {2},
        pages = {1434-1445},
          doi = {10.1086/523034},
archivePrefix = {arXiv},
       eprint = {0709.1720},
 primaryClass = {astro-ph},
       adsurl = {https://ui.adsabs.harvard.edu/abs/2007ApJ...671.1434T}
}

@ARTICLE{Tonnesen2008,
       author = {{Tonnesen}, Stephanie and {Bryan}, Greg L.},
        title = "{The Impact of ICM Substructure on Ram Pressure Stripping}",
      journal = {\apjl},
         year = 2008,
        month = sep,
       volume = {684},
       number = {1},
        pages = {L9},
          doi = {10.1086/592066},
archivePrefix = {arXiv},
       eprint = {0808.0007},
 primaryClass = {astro-ph},
       adsurl = {https://ui.adsabs.harvard.edu/abs/2008ApJ...684L...9T}
}

@ARTICLE{Tonnesen2009,
       author = {{Tonnesen}, Stephanie and {Bryan}, Greg L.},
        title = "{Gas Stripping in Simulated Galaxies with a Multiphase Interstellar Medium}",
      journal = {\apj},
         year = 2009,
        month = apr,
       volume = {694},
       number = {2},
        pages = {789-804},
          doi = {10.1088/0004-637X/694/2/789},
archivePrefix = {arXiv},
       eprint = {0901.2115},
 primaryClass = {astro-ph.GA},
       adsurl = {https://ui.adsabs.harvard.edu/abs/2009ApJ...694..789T}
}

@ARTICLE{Tonnesen2011,
       author = {{Tonnesen}, Stephanie and {Bryan}, Greg L. and {Chen}, Rena},
        title = "{How to Light it Up: Simulating Ram-pressure Stripped X-ray Bright Tails}",
      journal = {\apj},
         year = 2011,
        month = apr,
       volume = {731},
       number = {2},
          eid = {98},
        pages = {98},
          doi = {10.1088/0004-637X/731/2/98},
archivePrefix = {arXiv},
       eprint = {1103.3273},
 primaryClass = {astro-ph.CO},
       adsurl = {https://ui.adsabs.harvard.edu/abs/2011ApJ...731...98T}
}

@ARTICLE{Tonnesen2019,
       author = {{Tonnesen}, Stephanie},
        title = "{The Journey Counts: The Importance of Including Orbits when Simulating Ram Pressure Stripping}",
      journal = {\apj},
         year = 2019,
        month = apr,
       volume = {874},
       number = {2},
          eid = {161},
        pages = {161},
          doi = {10.3847/1538-4357/ab0960},
archivePrefix = {arXiv},
       eprint = {1903.08178},
 primaryClass = {astro-ph.GA},
       adsurl = {https://ui.adsabs.harvard.edu/abs/2019ApJ...874..161T}
}

@ARTICLE{Jachym2009,
       author = {{J{\'a}chym}, P. and {K{\"o}ppen}, J. and {Palou{\v{s}}}, J. and {Combes}, F.},
        title = "{Ram pressure stripping of tilted galaxies}",
      journal = {\aap},
         year = 2009,
        month = jun,
       volume = {500},
       number = {2},
        pages = {693-703},
          doi = {10.1051/0004-6361/200811469},
archivePrefix = {arXiv},
       eprint = {0904.3886},
 primaryClass = {astro-ph.CO},
       adsurl = {https://ui.adsabs.harvard.edu/abs/2009A&A...500..693J}
}

@ARTICLE{Jachym2007,
       author = {{J{\'a}chym}, P. and {Palou{\v{s}}}, J. and {K{\"o}ppen}, J. and {Combes}, F.},
        title = "{Gas stripping in galaxy clusters: a new SPH simulation approach}",
      journal = {\aap},
         year = 2007,
        month = sep,
       volume = {472},
       number = {1},
        pages = {5-20},
          doi = {10.1051/0004-6361:20066442},
archivePrefix = {arXiv},
       eprint = {0706.3631},
 primaryClass = {astro-ph},
       adsurl = {https://ui.adsabs.harvard.edu/abs/2007A&A...472....5J}
}

@ARTICLE{Kapferer2008,
       author = {{Kapferer}, W. and {Kronberger}, T. and {Ferrari}, C. and {Riser}, T. and {Schindler}, S.},
        title = "{On the influence of ram-pressure stripping on interacting galaxies in clusters}",
      journal = {\mnras},
         year = 2008,
        month = sep,
       volume = {389},
       number = {3},
        pages = {1405-1413},
          doi = {10.1111/j.1365-2966.2008.13665.x},
archivePrefix = {arXiv},
       eprint = {0807.0083},
 primaryClass = {astro-ph},
       adsurl = {https://ui.adsabs.harvard.edu/abs/2008MNRAS.389.1405K}
}

@ARTICLE{Kapferer2009,
       author = {{Kapferer}, W. and {Sluka}, C. and {Schindler}, S. and {Ferrari}, C. and {Ziegler}, B.},
        title = "{The effect of ram pressure on the star formation, mass distribution and morphology of galaxies}",
      journal = {\aap},
         year = 2009,
        month = may,
       volume = {499},
       number = {1},
        pages = {87-102},
          doi = {10.1051/0004-6361/200811551},
archivePrefix = {arXiv},
       eprint = {0903.3818},
 primaryClass = {astro-ph.CO},
       adsurl = {https://ui.adsabs.harvard.edu/abs/2009A&A...499...87K}
}

@ARTICLE{Yun2019,
       author = {{Yun}, Kiyun and {Pillepich}, Annalisa and {Zinger}, Elad and {Nelson}, Dylan and {Donnari}, Martina and {Joshi}, Gandhali and {Rodriguez-Gomez}, Vicente and {Genel}, Shy and {Weinberger}, Rainer and {Vogelsberger}, Mark and {Hernquist}, Lars},
        title = "{Jellyfish galaxies with the IllustrisTNG simulations - I. Gas-stripping phenomena in the full cosmological context}",
      journal = {\mnras},
         year = 2019,
        month = feb,
       volume = {483},
       number = {1},
        pages = {1042-1066},
          doi = {10.1093/mnras/sty3156},
archivePrefix = {arXiv},
       eprint = {1810.00005},
 primaryClass = {astro-ph.GA},
       adsurl = {https://ui.adsabs.harvard.edu/abs/2019MNRAS.483.1042Y}
}

@ARTICLE{Troncoso2020,
       author = {{Troncoso-Iribarren}, P. and {Padilla}, N. and {Santander}, C. and {Lagos}, C.~D.~P. and {Garc{\'\i}a-Lambas}, D. and {Rodr{\'\i}guez}, S. and {Contreras}, S.},
        title = "{The better half - asymmetric star formation due to ram pressure in the EAGLE simulations}",
      journal = {\mnras},
         year = 2020,
        month = oct,
       volume = {497},
       number = {4},
        pages = {4145-4161},
          doi = {10.1093/mnras/staa274},
archivePrefix = {arXiv},
       eprint = {2001.06501},
 primaryClass = {astro-ph.GA},
       adsurl = {https://ui.adsabs.harvard.edu/abs/2020MNRAS.497.4145T}
}

@ARTICLE{Joshi2020,
       author = {{Joshi}, Gandhali D. and {Pillepich}, Annalisa and {Nelson}, Dylan and {Marinacci}, Federico and {Springel}, Volker and {Rodriguez-Gomez}, Vicente and {Vogelsberger}, Mark and {Hernquist}, Lars},
        title = "{The fate of disc galaxies in IllustrisTNG clusters}",
      journal = {\mnras},
         year = 2020,
        month = aug,
       volume = {496},
       number = {3},
        pages = {2673-2703},
          doi = {10.1093/mnras/staa1668},
archivePrefix = {arXiv},
       eprint = {2004.01191},
 primaryClass = {astro-ph.GA},
       adsurl = {https://ui.adsabs.harvard.edu/abs/2020MNRAS.496.2673J}
}

@ARTICLE{Lee2020,
       author = {{Lee}, Jaehyun and {Kimm}, Taysun and {Katz}, Harley and {Rosdahl}, Joakim and {Devriendt}, Julien and {Slyz}, Adrianne},
        title = "{Dual Effects of Ram Pressure on Star Formation in Multiphase Disk Galaxies with Strong Stellar Feedback}",
      journal = {\apj},
         year = 2020,
        month = dec,
       volume = {905},
       number = {1},
          eid = {31},
        pages = {31},
          doi = {10.3847/1538-4357/abc3b8},
archivePrefix = {arXiv},
       eprint = {2010.11028},
 primaryClass = {astro-ph.GA},
       adsurl = {https://ui.adsabs.harvard.edu/abs/2020ApJ...905...31L}
}

@ARTICLE{Boselli2022,
       author = {{Boselli}, Alessandro and {Fossati}, Matteo and {Sun}, Ming},
        title = "{Ram pressure stripping in high-density environments}",
      journal = {\aapr},
         year = 2022,
        month = dec,
       volume = {30},
       number = {1},
          eid = {3},
        pages = {3},
          doi = {10.1007/s00159-022-00140-3},
archivePrefix = {arXiv},
       eprint = {2109.13614},
 primaryClass = {astro-ph.GA},
       adsurl = {https://ui.adsabs.harvard.edu/abs/2022A&ARv..30....3B}
}

@ARTICLE{Vollmer2021,
       author = {{Vollmer}, B. and {Fossati}, M. and {Boselli}, A. and {Soida}, M. and {Gwyn}, S. and {Cuillandre}, J.~C. and {Amram}, Ph. and {Boissier}, S. and {Boquien}, M. and {Hensler}, G.},
        title = "{A Virgo Environmental Survey Tracing Ionised Gas Emission (VESTIGE). VIII. Modeling ram pressure stripping of diffuse gas in the Virgo cluster spiral galaxy NGC 4330}",
      journal = {\aap},
         year = 2021,
        month = jan,
       volume = {645},
          eid = {A121},
        pages = {A121},
          doi = {10.1051/0004-6361/202038507},
archivePrefix = {arXiv},
       eprint = {2011.03437},
 primaryClass = {astro-ph.GA},
       adsurl = {https://ui.adsabs.harvard.edu/abs/2021A&A...645A.121V}
}

@ARTICLE{Churazov2023,
       author = {{Churazov}, E. and {Khabibullin}, I. and {Bykov}, A.~M. and {Lyskova}, N. and {Sunyaev}, R.},
        title = "{Tempestuous life beyond R$_{500}$: X-ray view on the Coma cluster with SRG/eROSITA. II. Shock and relic}",
      journal = {\aap},
         year = 2023,
        month = feb,
       volume = {670},
          eid = {A156},
        pages = {A156},
          doi = {10.1051/0004-6361/202244021},
archivePrefix = {arXiv},
       eprint = {2205.07511},
 primaryClass = {astro-ph.CO},
       adsurl = {https://ui.adsabs.harvard.edu/abs/2023A&A...670A.156C}
}

@ARTICLE{Owers2012,
       author = {{Owers}, Matt S. and {Couch}, Warrick J. and {Nulsen}, Paul E.~J. and {Randall}, Scott W.},
        title = "{Shocking Tails in the Major Merger Abell 2744}",
      journal = {\apjl},
         year = 2012,
        month = may,
       volume = {750},
       number = {1},
          eid = {L23},
        pages = {L23},
          doi = {10.1088/2041-8205/750/1/L23},
archivePrefix = {arXiv},
       eprint = {1204.1052},
 primaryClass = {astro-ph.CO},
       adsurl = {https://ui.adsabs.harvard.edu/abs/2012ApJ...750L..23O}
}

@ARTICLE{McPartland2016,
       author = {{McPartland}, Conor and {Ebeling}, Harald and {Roediger}, Elke and {Blumenthal}, Kelly},
        title = "{Jellyfish: the origin and distribution of extreme ram-pressure stripping events in massive galaxy clusters}",
      journal = {\mnras},
         year = 2016,
        month = jan,
       volume = {455},
       number = {3},
        pages = {2994-3008},
          doi = {10.1093/mnras/stv2508},
archivePrefix = {arXiv},
       eprint = {1511.00033},
 primaryClass = {astro-ph.GA},
       adsurl = {https://ui.adsabs.harvard.edu/abs/2016MNRAS.455.2994M}
}

@ARTICLE{Markevitch2001,
   author = {{Markevitch}, M. and {Vikhlinin}, A. and {Mazzotta}, P.},
    title = "{Nonhydrostatic Gas in the Core of the Relaxed Galaxy Cluster A1795}",
  journal = {\apjl},
   eprint = {arXiv:astro-ph/0108520},
     year = 2001,
    month = dec,
   volume = 562,
    pages = {L153-L156},
      doi = {10.1086/337973},
   adsurl = {http://adsabs.harvard.edu/abs/2001ApJ...562L.153M}
}

@ARTICLE{RomanOliveira2019,
       author = {{Roman-Oliveira}, Fernanda V. and {Chies-Santos}, Ana L. and {Rodr{\'\i}guez del Pino}, Bruno and {Arag{\'o}n-Salamanca}, A. and {Gray}, Meghan E. and {Bamford}, Steven P.},
        title = "{OMEGA-OSIRIS mapping of emission-line galaxies in A901/2-V. The rich population of jellyfish galaxies in the multicluster system Abell 901/2}",
      journal = {\mnras},
         year = 2019,
        month = mar,
       volume = {484},
       number = {1},
        pages = {892-905},
          doi = {10.1093/mnras/stz007},
archivePrefix = {arXiv},
       eprint = {1812.05629},
 primaryClass = {astro-ph.GA},
       adsurl = {https://ui.adsabs.harvard.edu/abs/2019MNRAS.484..892R}
}

@ARTICLE{Poggianti2016,
       author = {{Poggianti}, B.~M. and {Fasano}, G. and {Omizzolo}, A. and {Gullieuszik}, M. and {Bettoni}, D. and {Moretti}, A. and {Paccagnella}, A. and {Jaff{\'e}}, Y.~L. and {Vulcani}, B. and {Fritz}, J. and {Couch}, W. and {D'Onofrio}, M.},
        title = "{Jellyfish Galaxy Candidates at Low Redshift}",
      journal = {\aj},
         year = 2016,
        month = mar,
       volume = {151},
       number = {3},
          eid = {78},
        pages = {78},
          doi = {10.3847/0004-6256/151/3/78},
archivePrefix = {arXiv},
       eprint = {1504.07105},
 primaryClass = {astro-ph.GA},
       adsurl = {https://ui.adsabs.harvard.edu/abs/2016AJ....151...78P}
}

@ARTICLE{Dehnen1993,
       author = {{Dehnen}, W.},
        title = "{A Family of Potential-Density Pairs for Spherical Galaxies and Bulges}",
      journal = {\mnras},
         year = 1993,
        month = nov,
       volume = {265},
        pages = {250},
          doi = {10.1093/mnras/265.1.250},
       adsurl = {https://ui.adsabs.harvard.edu/abs/1993MNRAS.265..250D}
}

@ARTICLE{Lourenco2023,
       author = {{Louren{\c{c}}o}, Ana C.~C. and {Jaff{\'e}}, Y.~L. and {Vulcani}, B. and {Biviano}, A. and {Poggianti}, B. and {Moretti}, A. and {Kelkar}, K. and {Crossett}, J.~P. and {Gitti}, M. and {Smith}, R. and {Lagan{\'a}}, T.~F. and {Gullieuszik}, M. and {Ignesti}, A. and {McGee}, S. and {Wolter}, A. and {Sonkamble}, S. and {M{\"u}ller}, A.},
        title = "{The effect of cluster dynamical state on ram-pressure stripping}",
      journal = {\mnras},
         year = 2023,
        month = dec,
       volume = {526},
       number = {4},
        pages = {4831-4847},
          doi = {10.1093/mnras/stad2972},
archivePrefix = {arXiv},
       eprint = {2309.15934},
 primaryClass = {astro-ph.GA},
       adsurl = {https://ui.adsabs.harvard.edu/abs/2023MNRAS.526.4831L}
}

@ARTICLE{Goller2023,
       author = {{G{\"o}ller}, Junia and {Joshi}, Gandhali D. and {Rohr}, Eric and {Zinger}, Elad and {Pillepich}, Annalisa},
        title = "{Jellyfish galaxies with the IllustrisTNG simulations - No enhanced population-wide star formation according to TNG50}",
      journal = {\mnras},
         year = 2023,
        month = nov,
       volume = {525},
       number = {3},
        pages = {3551-3570},
          doi = {10.1093/mnras/stad2551},
archivePrefix = {arXiv},
       eprint = {2304.09199},
 primaryClass = {astro-ph.GA},
       adsurl = {https://ui.adsabs.harvard.edu/abs/2023MNRAS.525.3551G}
}

@ARTICLE{Vijayaraghavan2013,
       author = {{Vijayaraghavan}, R. and {Ricker}, P.~M.},
        title = "{Pre-processing and post-processing in group-cluster mergers}",
      journal = {\mnras},
         year = 2013,
        month = nov,
       volume = {435},
       number = {3},
        pages = {2713-2735},
          doi = {10.1093/mnras/stt1485},
archivePrefix = {arXiv},
       eprint = {1308.1311},
 primaryClass = {astro-ph.CO},
       adsurl = {https://ui.adsabs.harvard.edu/abs/2013MNRAS.435.2713V}
}

@ARTICLE{Conroy2010,
       author = {{Conroy}, Charlie and {Gunn}, James E.},
        title = "{The Propagation of Uncertainties in Stellar Population Synthesis Modeling. III. Model Calibration, Comparison, and Evaluation}",
      journal = {\apj},
         year = 2010,
        month = apr,
       volume = {712},
       number = {2},
        pages = {833-857},
          doi = {10.1088/0004-637X/712/2/833},
archivePrefix = {arXiv},
       eprint = {0911.3151},
 primaryClass = {astro-ph.CO},
       adsurl = {https://ui.adsabs.harvard.edu/abs/2010ApJ...712..833C}
}

@ARTICLE{Nelson2019,
       author = {{Nelson}, Dylan and {Springel}, Volker and {Pillepich}, Annalisa and {Rodriguez-Gomez}, Vicente and {Torrey}, Paul and {Genel}, Shy and {Vogelsberger}, Mark and {Pakmor}, Ruediger and {Marinacci}, Federico and {Weinberger}, Rainer and {Kelley}, Luke and {Lovell}, Mark and {Diemer}, Benedikt and {Hernquist}, Lars},
        title = "{The IllustrisTNG simulations: public data release}",
      journal = {Computational Astrophysics and Cosmology},
         year = 2019,
        month = may,
       volume = {6},
       number = {1},
          eid = {2},
        pages = {2},
          doi = {10.1186/s40668-019-0028-x},
archivePrefix = {arXiv},
       eprint = {1812.05609},
 primaryClass = {astro-ph.GA},
       adsurl = {https://ui.adsabs.harvard.edu/abs/2019ComAC...6....2N}
}

@ARTICLE{Albuquerque2024,
       author = {{Albuquerque}, Richards P. and {Machado}, Rubens E.~G. and {Monteiro-Oliveira}, Rog{\'e}rio},
        title = "{Unravelling the collision scenario of the dissociative galaxy cluster Abell 56 through hydrodynamic simulations}",
      journal = {\mnras},
         year = 2024,
        month = may,
       volume = {530},
       number = {2},
        pages = {2146-2155},
          doi = {10.1093/mnras/stae1004},
archivePrefix = {arXiv},
       eprint = {2401.15044},
 primaryClass = {astro-ph.GA},
       adsurl = {https://ui.adsabs.harvard.edu/abs/2024MNRAS.530.2146A}
}

@ARTICLE{Rohr2023,
       author = {{Rohr}, Eric and {Pillepich}, Annalisa and {Nelson}, Dylan and {Zinger}, Elad and {Joshi}, Gandhali D. and {Ayromlou}, Mohammadreza},
        title = "{Jellyfish galaxies with the IllustrisTNG simulations - when, where, and for how long does ram pressure stripping of cold gas occur?}",
      journal = {\mnras},
         year = 2023,
        month = sep,
       volume = {524},
       number = {3},
        pages = {3502-3525},
          doi = {10.1093/mnras/stad2101},
archivePrefix = {arXiv},
       eprint = {2304.09196},
 primaryClass = {astro-ph.GA},
       adsurl = {https://ui.adsabs.harvard.edu/abs/2023MNRAS.524.3502R}
}

@ARTICLE{Briel1992,
       author = {{Briel}, U.~G. and {Henry}, J.~P. and {Boehringer}, H.},
        title = "{Observation of the Coma cluster of galaxies with ROSAT during the all-sky-survey.}",
      journal = {\aap},
         year = 1992,
        month = jun,
       volume = {259},
        pages = {L31-L34},
       adsurl = {https://ui.adsabs.harvard.edu/abs/1992A&A...259L..31B}
}

@ARTICLE{Neumann2003,
       author = {{Neumann}, D.~M. and {Lumb}, D.~H. and {Pratt}, G.~W. and {Briel}, U.~G.},
        title = "{The dynamical state of the Coma cluster with XMM-Newton}",
      journal = {\aap},
         year = 2003,
        month = mar,
       volume = {400},
        pages = {811-821},
          doi = {10.1051/0004-6361:20021911},
archivePrefix = {arXiv},
       eprint = {astro-ph/0212432},
 primaryClass = {astro-ph},
       adsurl = {https://ui.adsabs.harvard.edu/abs/2003A&A...400..811N}
}

@INPROCEEDINGS{Biviano1998,
       author = {{Biviano}, A.},
        title = "{Our best friend, the Coma cluster (a historical review)}",
    booktitle = {Untangling Coma Berenices: A New Vision of an Old Cluster},
         year = 1998,
       editor = {{Mazure}, A. and {Casoli}, F. and {Durret}, F. and {Gerbal}, D.},
        month = jan,
        pages = {1},
          doi = {10.48550/arXiv.astro-ph/9711251},
archivePrefix = {arXiv},
       eprint = {astro-ph/9711251},
 primaryClass = {astro-ph},
       adsurl = {https://ui.adsabs.harvard.edu/abs/1998ucb..proc....1B}
}

@ARTICLE{Colless1996,
       author = {{Colless}, Matthew and {Dunn}, Andrew M.},
        title = "{Structure and Dynamics of the Coma Cluster}",
      journal = {\apj},
         year = 1996,
        month = feb,
       volume = {458},
        pages = {435},
          doi = {10.1086/176827},
archivePrefix = {arXiv},
       eprint = {astro-ph/9508070},
 primaryClass = {astro-ph},
       adsurl = {https://ui.adsabs.harvard.edu/abs/1996ApJ...458..435C}
}

\end{document}